\documentclass[prl,twocolumn]{revtex4}
\usepackage{amssymb}
\usepackage{amsmath,bbm}
\usepackage{epsfig}
\usepackage{enumerate}
\usepackage{subfigure}
\newcommand{\be}{\begin{equation}}
\newcommand{\ee}{\end{equation}}
\newcommand{\bea}{\begin{eqnarray}}
\newcommand{\eea}{\end{eqnarray}}

\newcommand{\cN}{{\cal N}}
\newcommand{\bi}{\begin{itemize}}
\newcommand{\ei}{\end{itemize}}
\newcommand{\ii}{\item}
\newcommand{\qf}{{\cal{Q}}}
\begin{document}
\title{Systematics of geometric scaling}
\author{F. Gelis}\email{gelis@spht.saclay.cea.fr}
\affiliation{Service de physique th{\'e}orique, CEA/Saclay, 91191 
Gif-sur-Yvette 
cedex, France\\URA 2306, unit{\'e} de recherche associ{\'e}e au CNRS}
\author{R. Peschanski}\email{pesch@spht.saclay.cea.fr}
\affiliation{Service de physique th{\'e}orique, CEA/Saclay, 91191 
Gif-sur-Yvette 
cedex, France\\URA 2306, unit{\'e} de recherche associ{\'e}e au CNRS}
\author{G. Soyez\footnote{on leave from the PTF group of the University of 
Li\`ege. Funded by the National Funds for Scientific Research (FNRS, Belgium)}}
\email{g.soyez@ulg.ac.be}
\affiliation{Service de physique th{\'e}orique, CEA/Saclay, 91191 
Gif-sur-Yvette 
cedex, France\\URA 2306, unit{\'e} de recherche associ{\'e}e au CNRS}
\author{L. Schoeffel}\email{schoeffel@hep.saclay.cea.fr}
\affiliation{DAPNIA/Service de physique des particules, CEA/Saclay, 91191 
Gif-sur-Yvette cedex, France}
%%%%%%%%%%%%%%%%%%%%%%%%%%%%%%%%%%%%%%%%%%%%%%%%%%%%%%%%%%%%%%%%%%
%%%%%%%%%%%%%%%%%%%%%%%%   Abstract   %%%%%%%%%%%%%%%%%%%%%%%%%%%%
%%%%%%%%%%%%%%%%%%%%%%%%%%%%%%%%%%%%%%%%%%%%%%%%%%%%%%%%%%%%%%%%%%
\begin{abstract}

Using all available data on the deep-inelastic cross-sections at HERA at $x\le 
10^{-2}$, we look for geometric scaling of the form $\sigma 
^{\gamma^*p}(\tau)$ where the scaling variable $\tau$ behaves alternatively 
like 
$\log Q^2\!-\!\lambda Y,$ as in the original definition, or $\log Q^2\!-\!\lambda 
\sqrt 
Y,$ which is  suggested by the asymptotic properties of the 
Balitsky-Kovchegov (BK) equation with {\it running} QCD coupling constant. A 
``Quality Factor'' (QF) is defined, quantifying  the phenomenological validity of 
the scaling and 
the uncertainty on  the intercept $\lambda.$ Both choices have a good QF, showing 
that the second choice is as valid as the first one, predicted for {\it fixed} 
coupling constant. A comparison between the QCD asymptotic predictions and data is 
made and the QF analysis shows that the agreement can be reached, provided  going 
beyond leading logarithmic accuracy for the BK equation. 

\end{abstract}
\maketitle

%%%%%%%%%%%%%%%%%%%%%%%%%%%%%%%%%%%%%%%%%%%%%%%%%%%%%%%%%%%%%%%%%%
%%%%%%%%%%%%%%%%%%%%%%%%   Section 1   %%%%%%%%%%%%%%%%%%%%%%%%%%%
%%%%%%%%%%%%%%%%%%%%%%%%%%%%%%%%%%%%%%%%%%%%%%%%%%%%%%%%%%%%%%%%%%

{\bf 1.} Geometric scaling \cite{Stasto:2000er} is a remarkable empirical 
property verified by the data on the high-energy deep-inelastic scattering (DIS) 
cross-sections $\sigma^{\gamma^*p}.$ It has been realized that one can represent 
with reasonable accuracy the cross-section by the formula
\begin{equation}
\sigma^{\gamma^*p}(Y,Q)=\sigma^{\gamma^*p}\!
\left(\frac{Q^2}{Q_s^2(Y)}\right)\ ,
\label{geometric}
\end{equation}
where $Q$ is the virtuality of the photon, $Y$ the total rapidity in the 
${\gamma^*}$-proton system and $Q_s^2 \propto e^{\lambda Y}$ an increasing function 
of $Y$. The value found for  $\lambda \sim 0.3$ has been confirmed by 
the well-known Golec-Biernat and W\"usthoff model \cite{Golec-Biernat:1998js} 
where geometric scaling has been explicitly assumed in the parametrisation.

The property \eqref{geometric}, also observed in DIS on nucleus \cite{adis} and
diffractive processes \cite{ms}, has been intimately related \cite{iim} to the 
concept of {\it saturation} \cite{saturation}, i.e. the behaviour of perturbative 
QCD amplitudes when the density of partons becomes high enough to exercise the 
unitarity bound. Indeed, there has been many theoretical arguments  to 
infer that in a $Y, Q^2$ domain  where  saturation effects  set in, geometric 
scaling is expected to occur. Within this framework, the function  $Q_s (Y)$ can be 
called the saturation scale, since it  delineates the approximate upper bound of the 
saturation domain.

Following  a  theoretical approach, it has been possible to derive the 
property \eqref{geometric} 
from the  nonlinear Balitsky-Kovchegov 
(BK) equation which represents the ``mean-field'' approximation of the evolution equation 
for high energy (high 
density) QCD amplitudes. This equation is supposed to capture essential features of 
saturation effects. In the 1-dimensional  approximation it reads for the density 
of gluons with transverse 
momenta $k$ in some target \cite{Balitsky:1995ub}
\be
\partial_Y \cN (L, Y) = 
\bar\alpha\, \chi (-\partial_L)\: \cN (L, Y) - \bar\alpha\, \cN^{\,2} (L, Y),
\label{1}\ee
where $L \!= \!\log (k^2/k^2_0)$, with $k^2_0$ being an arbitrary constant. In 
Eq.\eqref{1}, the coupling constant will be considered alternatively in the 
following as fixed (we shall use $\bar\alpha=0.15$), or running 
such that $\bar\alpha(L)\! = \!{1}/{bL}\ ,
b = (11 N_c \!- \!2 N_f)/{12 N_c}.$ The kernel will also be either taken with 
leading logarithm (LL) accuracy \cite{Lipatov:1976zz}, as $
\chi(\gamma) \!=\! 2\psi(1) \!-\! \psi(\gamma)\! -\! \psi(1\!-\!\gamma)\ ,$
or at next-to-leading logarithm (NLL) accuracy (see \emph{e.g.} 
\cite{Fadin:1998py,Salam:1998tj,Ciafaloni:1999yw}).

The key ingredient to theoretically prove  geometric scaling of the asymptotic
solutions of the nonlinear equation \eqref{1} is the {\it travelling wave} 
method \cite{Munier:2003vc}. Indeed, the BK equation admits 
solutions in the form of {\it travelling waves} $\cN (L\! -\!\upsilon_g \bar\alpha 
Y)$. $L$ has the interpretation of a space variable while 
$t \!= \!\bar \alpha Y$, interpreted as time, is an increasing function of 
rapidity $Y.$  $\upsilon_g$ is the {\it critical velocity} of the wave, 
defined in 
this case as the minimum of the phase velocity. 
The travelling-wave solution for the quantity 
$\cN$ can be easily translated to the property \eqref{1} (at least assuming 
negligible quark masses) since it  yields 
$\cN (k^2/Q_s^2(Y))$. Hence the asymptotic travelling-wave solutions of the BK 
equation 
satisfy 
geometric scaling (up to subdominant scaling violations which we will not 
analyse in the present work). Similar results can be obtained 
\cite{Mueller:2002zm, dnt} from an approximation of the 
linear equation with absorptive boundary conditions.
\begin{figure*}[ht]
\subfigure [\ Quality Factor as a function of $\lambda.$]
{\epsfig{file=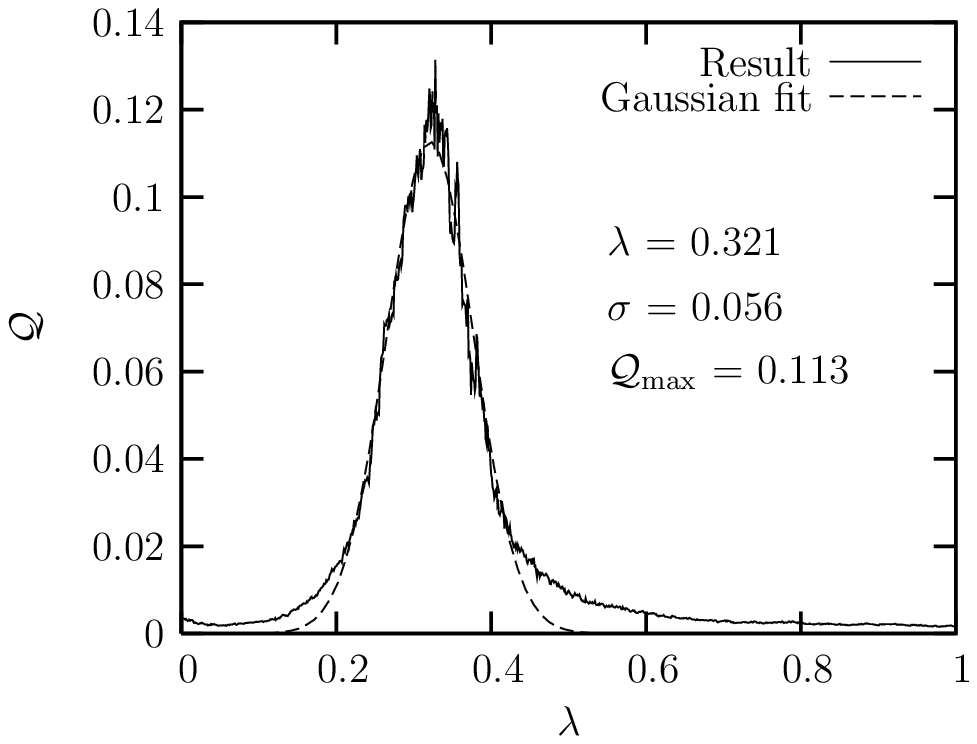,width=7.0cm}}\ \ \ \ \ \ \ \ \ \ \
\subfigure [\ Scaling~curve.] 
{\epsfig{file=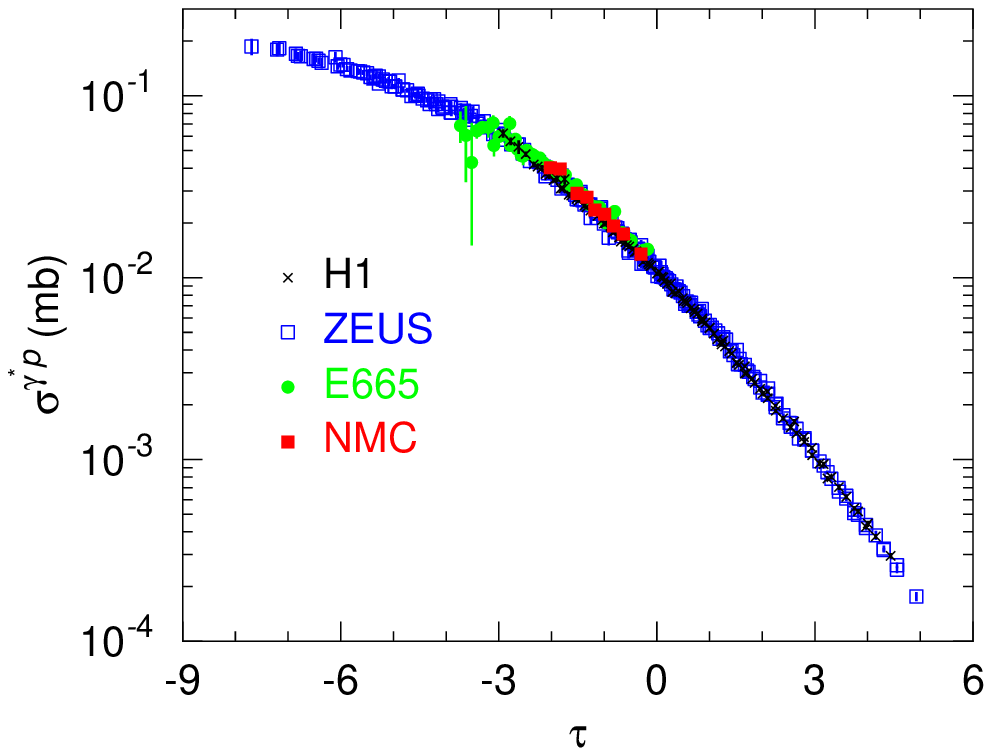,width=7.0cm}} 
\caption{Geometric Scaling in $Y$}\label{fix_Y}
\end{figure*}
The  main theoretical results for the asymptotic saturation scale are the 
following:
\vspace{-.5cm}
\begin{multline}
\log Q_s^2
=\bar\alpha\frac{\chi(\gamma_c)}{\gamma_c} (Y-Y_0)-\frac{3}{2\gamma_c}\log 
(Y-Y_0) \\
-\frac{3}{\gamma_c^2}
\sqrt{\frac{2\pi}{\bar\alpha\chi^{\prime\prime}(\gamma_c)}}\frac{1}{\sqrt{(Y+Y
_0)}}+...\ ,
\label{satscal}
\end{multline}
for  fixed coupling and
\begin{multline}
\log Q_s^2=\sqrt{\frac{2\chi(\gamma_c)}{b\gamma_c}(Y-Y_0)}
\\+{\scriptstyle \frac34}\left(\frac{\chi ^{\prime\prime}(\gamma_c)}
{\sqrt{2b\gamma_c\chi(\gamma_c)}}
\right)^{1/3}\!\!\!\!\!\!\!\xi_1\ (Y-Y_0)^{1/6}+...\ ,
\label{qsrunningalpha}
\end{multline} 
for  running coupling. $\gamma_c$ is the solution of the implicit equation
$\chi(\gamma_c)=\gamma_c\chi^\prime(\gamma_c)$ while $\xi_1=-2.338$ is the first 
zero of the Airy function. $Y_0$ is an arbitrary constant which may 
parameterize the unknown ``non-universal'' preasymptotic contributions, depending 
on the initial conditions.

The questions we want to ask deal with the 
confrontation between the empirical formula \eqref{1} with the asymptotic 
predictions for the saturation scale (\ref{satscal},\ref{qsrunningalpha}). 
They may be expressed as follows
\bi
\ii
Is geometric scaling, demonstrated for $\log(Q_s^2) \sim 
Y$ in agreement 
with the leading term in \eqref{satscal}, remains valid for $\log(Q_s^2) \sim 
Y^{1/2}$ as suggested by
\eqref{qsrunningalpha}?
\ii
Are the theoretical factors and 
subasymptotic terms given  in Eqs.(\ref{satscal},\ref{qsrunningalpha}) visible
in the data?
\ii
Are   NLL effects in the theoretical predictions phenomenologically 
important?
\ii
What is the role of ``non-universal''  terms?
\ei

{\bf 2.} We focus our analysis on the measurements of the  $\gamma^*p$ 
cross-section 
$\sigma^{\gamma^*p}=(4\pi^2\alpha_e/Q^2) F_2$ for which geometric scaling is 
predicted. The latter being valid at high-energy, we shall restrict ourselves 
to $x \le 0.01$ for all values of $Q^2$. Within that range, we shall use all 
available data (404 points from E665 \cite{E665}, H1 \cite{H1}, NMC \cite{NMC} and 
ZEUS \cite{ZEUS}).

In order to get a quantitative answer, we shall 
introduce an ``estimator'' or  {\it quality factor} (QF) for determining the 
scaling quality, whose working 
definition is the following. 
Given a set of points $(x_i,y_i,f_i\equiv f(x_i,y_i))$ and a parametric {\em 
scaling law} $\tau(x,y;\lambda)$, we want to determine if for some values of 
$\lambda$, $f(x,y)$ can be considered a function of $\tau$ only. To achieve 
this we define a QF  which is large when the points 
$(\tau_i=\tau(x_i,y_i;\lambda), f_i)$ ``lies on a unique curve''. To quantify 
this still ill-defined concept, we shall first rescale the set 
$(\tau_i,f_i)$ into $(u_i,v_i)$ such that $0\le u_i,v_i\le 1$ and assume that 
the $u_i$'s are ordered. We then introduce
\begin{equation}\label{eq:qf}
\qf(\lambda) = \left[\sum_i 
\frac{(v_i-v_{i-1})^2}{(u_i-u_{i-1})^2+\varepsilon^2}\right]^{-1}.
\end{equation}
This definition for the quality factor obviously achieves what we want: 
when two successive points are close in $u$ and far in $v$, we expect them 
``not to lie on the same curve'' and, indeed, they give a large contribution 
to the sum in \eqref{eq:qf}, leading to a small quality factor. The constant 
$\varepsilon^2$ in \eqref{eq:qf} is a small number (we have taken $\varepsilon=1/n$ 
with $n$ being the number of data points) which prevents the sum from becoming 
infinite when two points have the same value for $u$. Finally, by studying the 
dependence on $\lambda$ of $\qf$ (as we shall see, it usually shows a Gaussian 
peak) we can determine the best choice for the parameter in the scaling 
function $\tau$ and its uncertainty.
\begin{figure*}[ht] 
\subfigure [\ Quality Factor as a function of $\lambda.$]
{\epsfig{file=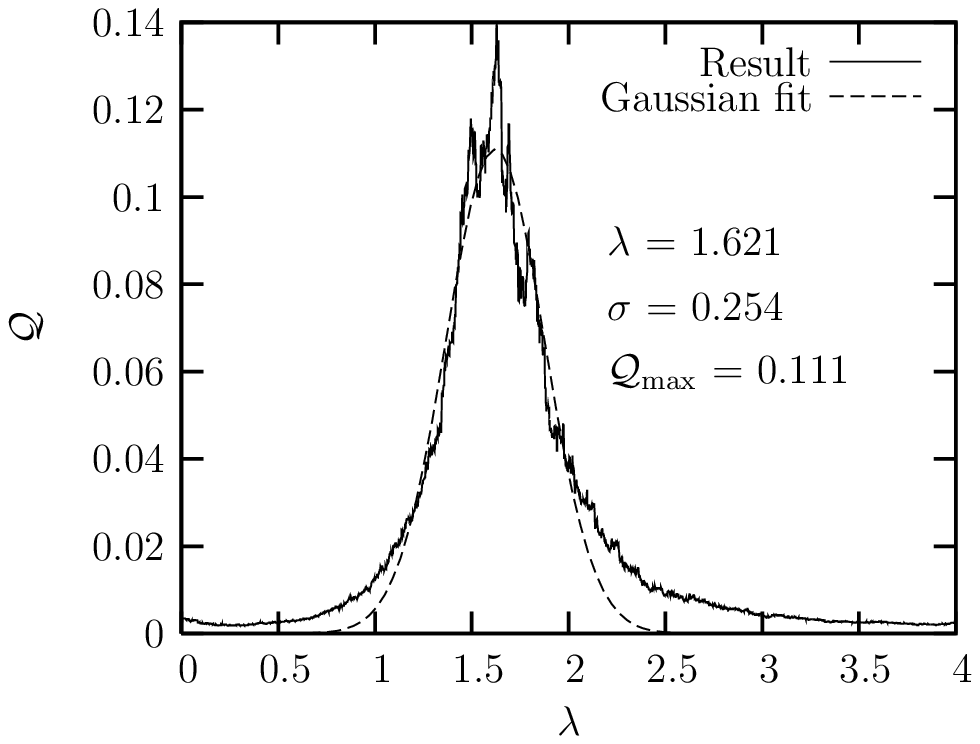,width=7.0cm}}\ \ \ \ \ \ \ \ \ \ \
\subfigure [\ Scaling~curve.] 
{\epsfig{file=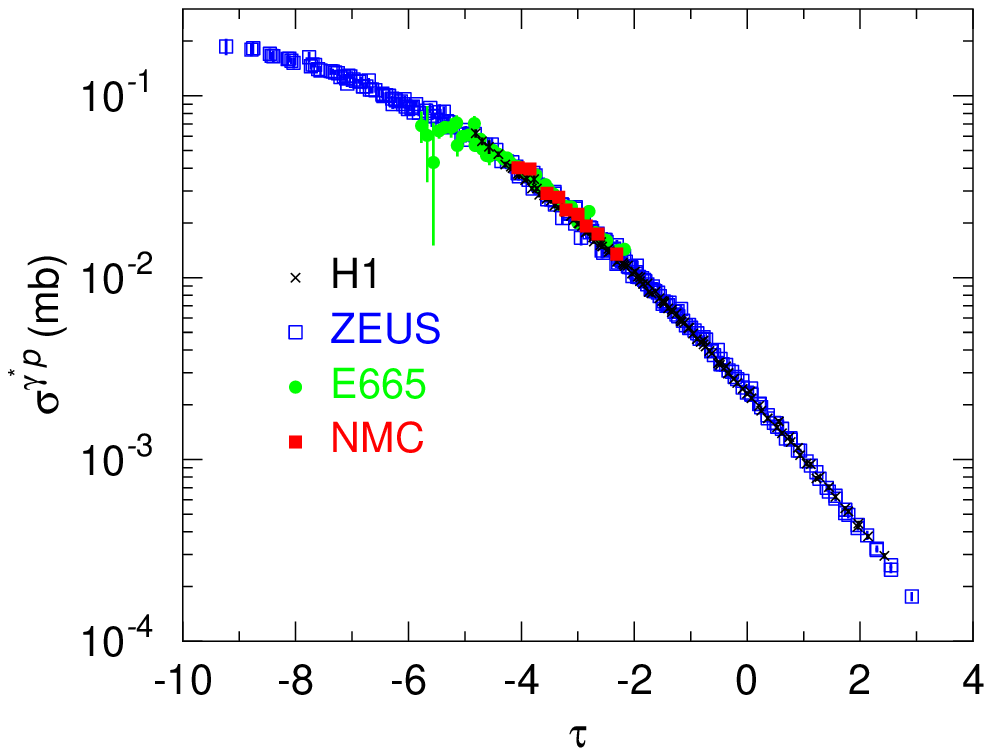,width=7.0cm}}
\caption{Geometric Scaling in $\sqrt Y$}\label{run_Y}
\end{figure*}

{\bf 3.} In order to answer the first question concerning the compared validity 
of geometric scaling in $Y$ \emph {vs.} $\sqrt Y,$ we performed the QF analysis 
in both cases. In Fig.\ref{fix_Y}, we show the data and QF for the usual 
geometric scaling definition with $Y.$ In Fig.\ref{fix_Y}-(b), is displayed  the  
data plot with scale redefinition. The quality of scaling on this plot can be 
read from  Fig.\ref{fix_Y}-(a), where the scatter plot of the QF is displayed 
together with a Gaussian fit of the peak. As obvious from Fig.\ref{fix_Y}-(b), 
the peak value larger than $0.1$ 
showss a good scaling and a Gaussian fit of the bump gives the best value and 
error for the lambda parameter $\lambda = 0.321 \pm 0.056\ .$ Note that the 
results displayed on  Fig.\ref{fix_Y} can be considered as an update for the 
usual geometric scaling tests and a quantitative determination of the scaling 
parameter and its attached uncertainty.
The new point is that geometric scaling is also verified, and with the same 
level of quality than previously, when the saturation scale is chosen as a 
function of $\sqrt Y.$ This is manifest on Fig.\ref{run_Y}, where comparable QF 
heights at the peak and width attest of the quality of scaling. The peak value 
getting larger than $0.11$ givess a good scaling. The Gaussian fit of the bump 
gives  $\lambda = 1.621 \pm 0.254\ .$
\begin{figure}[ht]
\begin{center}
{\epsfig{file=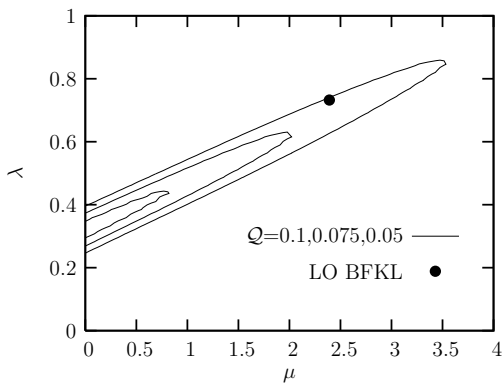,width=6.5cm}}
\caption{Quality Factor for the LL prediction.}\label{lambmu}
\end{center}
\end{figure}

{\bf 4.} Having verified  geometric scaling, as  
suggested by the asymptotic solutions of the QCD equations with both fixed and 
running coupling constant, it is tempting to go one step further 
and to ask  whether the theoretical formulae  
(\ref{satscal},\ref{qsrunningalpha}) with their theoretically predicted 
parametrisations are valid?

For that sake, we have considered first the LL prediction with fixed coupling 
\eqref{satscal} with the first subleading correction, {\em i.e.} 
of the  form $\log (Q_s^2) \propto \lambda Y \!- \!\mu \log(Y)$ (see Fig. 
\ref{lambmu}). The predicted values from \eqref{satscal} are represented by a 
point in the $\lambda,\mu$ plane. It happens that  a QF of $\sim 0.055$ can be 
attributed to this parametrisation, as shown on the same figure by the equi-QF  
curves. In this case changing $Y_0$ or adding the third term of \eqref{satscal}
does not improve the QF's. Hence the asymptotic LL theoretical prediction remains 
only  marginally verified at present energies.

%On Fig.\ref{NLL} we show the resulting analysis for 
Let us now turn to the analysis of
the 3 different theoretical 
NLL schemes that we  consider. 
\begin{figure*}[ht]
\subfigure [\ Quality Factor as a function of $Y_0$]
{\epsfig{file=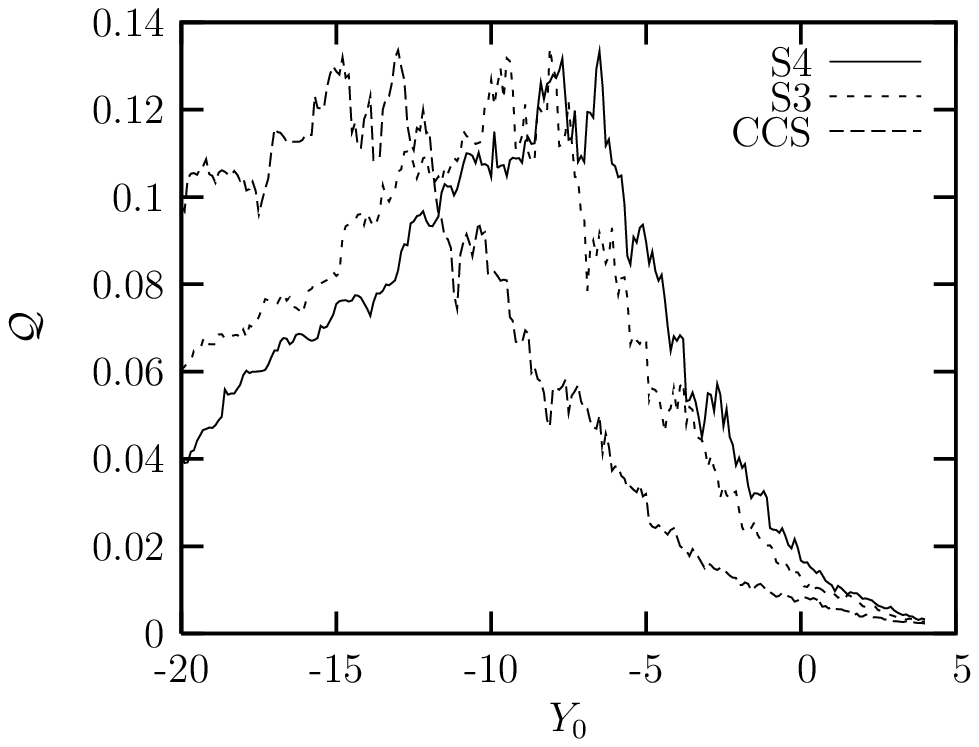,width=7.0cm}}\ \ \ \ \ \ \ \ \ \ \
\subfigure [\ Scaling~curve for the S4 scheme.] 
{\epsfig{file=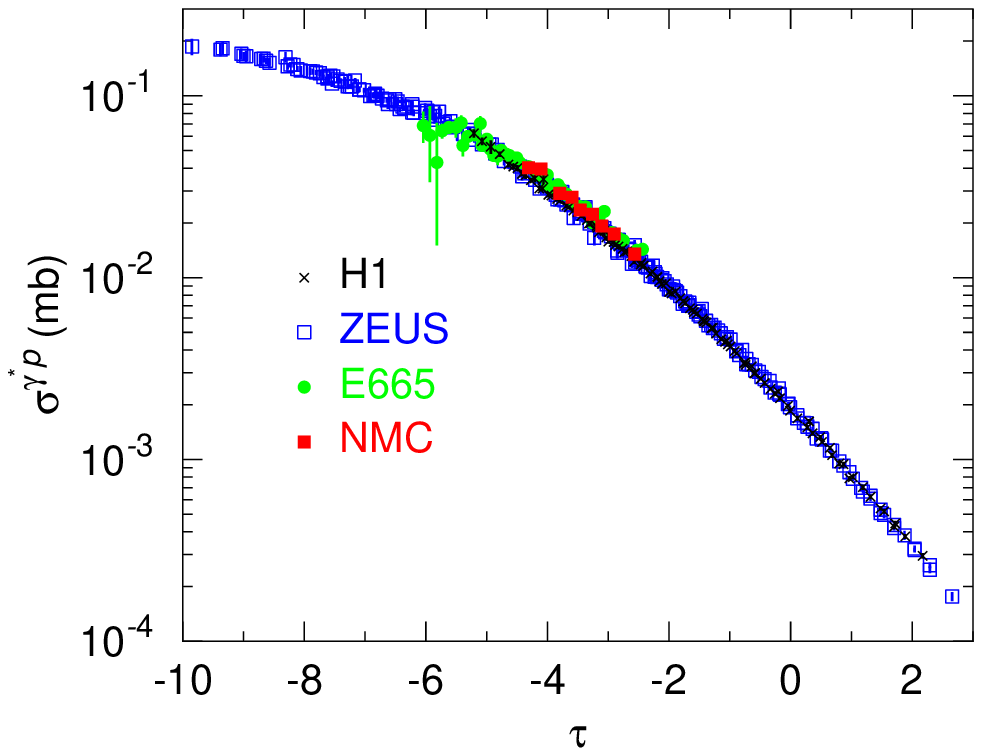,width=7.0cm}} 
\caption{Geometric Scaling at NLL accuracy}\label{NLL}
\end{figure*}
As now well known, NLL corrections to the LL kernel \cite{Fadin:1998py} have to be 
embedded in a resummation scheme in order to cancel 
spurious singularities.  We consider asymptotic predictions \eqref{qsrunningalpha} 
for the so-called S3, S4 schemes  \cite{Salam:1998tj} and  CCS 
scheme \cite{Ciafaloni:1999yw} which were recently derived \cite{sapeta}.

On Fig.\ref{NLL} we show the resulting analysis. 
The Quality Factor for the 3 schemes is presented as a function of $Y_0$, which 
parameterizes in \eqref{qsrunningalpha} the non-universal contributions. There 
always exists a  range of $Y_0$ for which the QF goes beyond 0.1 and thus leads to 
an acceptable scaling. This has to be contrasted with the situation with LL 
kernel. However, if  $Y_0$ is too negative, the relevance of geometrical scaling 
in $\sqrt Y$ is questionable. Indeed  $t = \sqrt{Y-Y_0} \sim \sqrt{\vert Y_0\vert} 
+  Y/{2\sqrt{\vert Y_0\vert}}$ and thus the scaling variable is $Y$ with a non 
universal $\lambda=1/{2\sqrt{\vert Y_0\vert}}$ parameter. For that reason the S4 
scheme seems favoured by the analysis and the figure displays the resulting scaling 
curve for  $Y_0 =-5.5.$

{\bf 5.} Using the preceding results based on the estimate of Quality Factors for 
scaling properties, it is now possible to answer the questions asked in the 
introduction. %Let us consider them in turn.
\bi
\ii
Geometric scaling is definitely as valid for the choice $\log(Q_s^2) \sim 
Y^{1/2}$ as it is for the original $\log(Q_s^2) \sim Y$ suggestion
\ii
The theoretical predictions \eqref{satscal} based on the BK equation at leading 
log accuracy, \emph{i.e.} with fixed coupling and LL kernel are only marginally 
verified with a QF of $\sim 0.055.$
\ii
At next-to-leading accuracy, the theoretical predictions  \eqref{qsrunningalpha} 
on scaling may be satisfied. There exists preferable NLL schemes \emph{e.g.} the S4 
scheme \cite{Salam:1998tj}.
\ii
Parametrising ``non-universal'' terms, \emph{i.e.} depending on initial conditions 
by a constant $Y_0$ in the rapidity evolution, we find that they do not play an 
essential role   at LL level, while they contribute to the quality of the 
scaling  at NLL level. 
\ei

The generality of the Quality Factor method indicates its interest for a 
quantitative evaluation of good scaling properties beyond the specific 
applications we focussed on in the present work. It can be used to check empirical 
scaling-law proposals as well as to quantitatively evaluate the ``distance'' 
between given theoretical predictions with data. In the present application, it 
leads to the conclusion that geometrical scaling is a well verified empirical 
property of deep-inelastic cross-sections on the proton. It also shows that the 
theoretical predictions based on the QCD saturation mechanism are consistent with 
the observation, but requires to go beyond leading logarithmic accuracy and the 
asymptotic terms.

Finally, let us quote that recent high-energy predictions \cite{himst} of {\em diffusive scaling}, 
namely $\sigma^{\gamma^*p}/\sqrt{Y} = \sigma(\tau)$ with $\tau=\log(Q^2/Q_s^2)/\sqrt{Y}$,
have a QF of 0.05, indicating that it might require higher energies.

%%%%%%%%%%%%%%%%%%%%%%%%%%%%%%%%%%%%%%%%%%%%%%%%%%%%%%%%%%%%%%%%%%
%%%%%%%%%%%%%%%%%%%%%%%%  Inter - Section  %%%%%%%%%%%%%%%%%%%%%%%%%%%
%%%%%%%%%%%%%%%%%%%%%%%%%%%%%%%%%%%%%%%%%%%%%%%%%%%%%%%%%%%%%%%%%%

%%%%%%%%%%%%%%%%%%%%%%%%%%%%%%%%%%%%%%%%%%%%%%%%%%%%%%%%%%%%%%%%%%
%%%%%%%%%%%%%%%%%%%%%%   Bibliography   %%%%%%%%%%%%%%%%%%%%%%%%%%
%%%%%%%%%%%%%%%%%%%%%%%%%%%%%%%%%%%%%%%%%%%%%%%%%%%%%%%%%%%%%%%%%%

\end{document}